\documentstyle[aas2pp4]{article}

\begin{document}

\title{Strong Correlation Between Noise Features at Low Frequency and the
Kilohertz QPOs in the X-Ray Binary 4U~1728--34}

\author{Eric C. Ford and Michiel van der Klis}

\authoremail{ecford@astro.uva.nl}
\affil{Astronomical Institute, ``Anton Pannekoek'',
University of Amsterdam, Kruislaan 403, 1098 SJ Amsterdam,
The Netherlands}

\vspace{0.5in}
\centerline {\bf ASTROPHYSICAL JOURNAL LETTERS: IN PRESS}

\begin{abstract}

We study the timing properties of the low mass X-ray binary 4U~1728--34
using recently released data from the Rossi X-Ray Timing
Explorer. This binary, like many others with accreting neutron stars,
is known to exhibit strong quasi-periodic oscillations (QPOs) of its
X-ray flux near 1 kHz. In addition to the kilohertz QPOs, the Fourier
power spectra show a broken power law noise component, with a break
frequency between 1 and 50 Hz, and a Lorentzian between 10 and 50
Hz. We find that the frequencies of the break and the low--frequency
Lorentzian are well correlated with the frequencies of the kilohertz
QPOs. The correlation between the frequency of the low--frequency
Lorentzian and the kilohertz QPO follows a power law relation with an
index of $2.11\pm 0.06$. The Lorentzian feature has been interpreted
as an effect of relativistic frame dragging (Lense--Thirring
precession) in the inner accretion disk (Stella \& Vietri 1998). This
model predicts a scaling index of 2, close to the value we measure.
This correlation is nearly identical to the one found in Z--sources
between the the well known QPOs on the horizontal branch and the
kilohertz QPOs, suggesting that the low frequency oscillations are a
similar phenomenon in these sources.  The frequency of the break in
the power spectra is also correlated with the frequencies of the
kilohertz QPOs. As previously noted for the similar binaries
4U~1608-50 and 4U~1705-44, this broken power law component closely
resembles that of black hole candidates in the low state, where the
break frequency is taken as a indicator of mass accretion rate. The
relation between break frequency and kilohertz QPO frequency thus
provides additional proof that the frequency of the kilohertz QPOs
increases with mass accretion rate.

\end{abstract}


\section{Introduction}

In most well observed low mass X-ray binaries, oscillations of the
X-ray flux at frequencies near 1 kHz have been measured with Rossi
X-ray Timing Explorer (for reviews and references see van der Klis
1998, Swank 1998). Models describing the production of these
quasi-periodic oscillations (QPOs) are not yet definitive, but the
fast signals most likely originate in the interesting region of the
accretion disk very close to the compact object where the dynamical
timescale is short.

The goal of this paper is to explore the link between the kilohertz
QPOs and the broadband timing properties in an X--ray binary. These
timing signals may well be related.  A common dependence on mass
accretion rate is suggested, since the strength of the low frequency
noise decreases with accretion rate while the frequency of the
kilohertz QPOs is expected to increase in current models.  The timing
features at tens of hertz may also be related to the kilohertz QPOs,
as in the Lense--Thirring precession model of Stella \& Vietri (1997)
or the oscillation mode model of Titarchuk, Lapidus \& Muslimov
(1998).

For this analysis we use RXTE observations of 4U~1728--34 (GX~354+0)
which have recently become public. Analysis of the kilohertz QPOs in
these observations was reported by Strohmayer et al. (1996). Section~2
summarizes the observations and our analysis of the Fourier power
spectra. In Section~3 we present the correlations between the
frequencies of the noise components. Section~4 is a discussion of the
results and the physical implications.

\section{Observations \& Analysis}

The present RXTE observations of 4U~1728--34 took place between 1996
February 15 and March 1 and consist of approximately 298 ksec of
usable data. We use here data from the proportional counter array
(PCA; for more instrument information see Zhang et al. 1993). We
generate Fourier power spectra from the high time resolution data,
which consists variously of a `single-bit' or an `event' mode. The
time resolutions in these modes range from 1.2 to 122 $\mu$sec; we use
a maximum Nyquist frequency of 4096~Hz. We employ all the channels of
the PCA, most sensitive in the 2--30 keV band.  There are 12 X-ray
bursts (Strohmayer, Zhang \& Swank 1997). We exclude these from our
analysis, excising 300 sec of data after each burst.

\begin{figure*}
\figurenum{1}
\epsscale{1.9}
\plotone{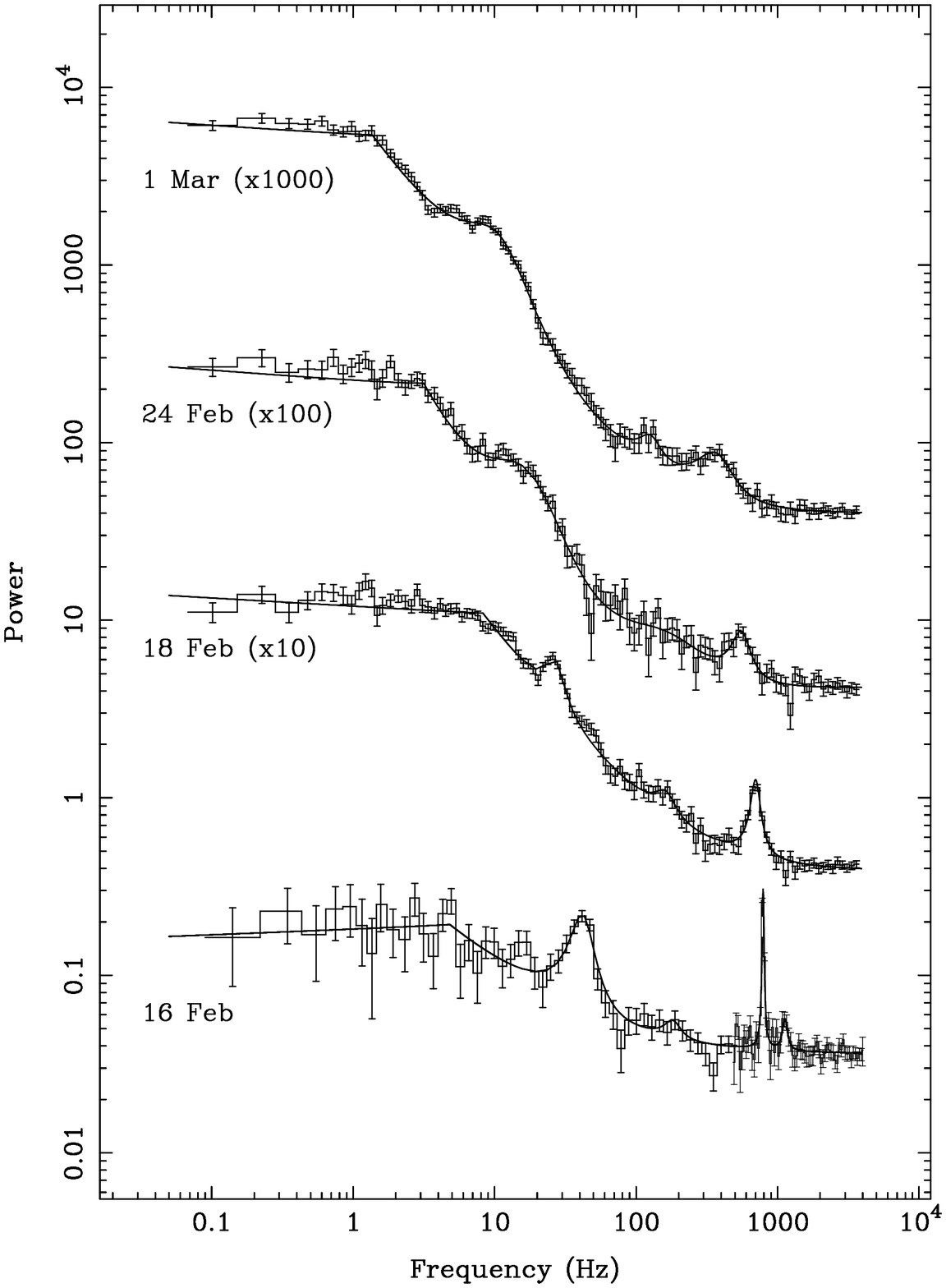} 
\caption{Representative power density spectra for four
observations. The fit parameters are listed in Table~1. The power is
Leahy normalized. We subtract a constant value of 1.95, somewhat less
than the Poisson level, and multiply each spectrum by a constant as
indicated. Fit functions are shown which consist of a broken power
law, a low frequency Lorentzian, a Lorentzian near 100 Hz, and
kilohertz QPO(s).}
\label{fig:pds}
\end{figure*}

Four representative power spectra are shown in Figure~\ref{fig:pds}.
We have arrived at a simple function that describes all the power
spectra well. The components are ({\it i}) a broken power law with a
break frequency whose best--fit value ranges between 1 and 50 Hz,
({\it ii}) the kilohertz QPOs, described by Lorentzians with best--fit
centroid values from 325 to 1127~Hz and ({\it iii}) a low--frequency
Lorentzian with a best--fit frequency from 10 to 50 Hz and $FWHM$
(full width at half maximum) between 3 and 30 Hz. In addition to these
three components we find excess power near 100 Hz. We include this in
the fit as ({\it iv}) another Lorentzian, with a best--fit frequency
of about 150 Hz ($FWHM$: 10--300 Hz). Including the fourth
component does not significantly change the best--fit values of the
other components.  All the parameters are allowed to vary
independently, and the reduced $\chi^2$ for the fits are typically 1.0
to 1.3 with about 200 degrees of freedom.  Table~1 shows the
parameters for the fits to the power spectra in
Figure~\ref{fig:pds}. Other atoll sources also show these same
components in their power spectra, as seen with Ginga observations of
4U~1608-52 (Yoshida et al. 1993) and RXTE observations of 4U~1705-44
(Ford et al. 1998), 4U0614+091 (Ford 1997, Mendez et al. 1997, van
Straaten et al. 1998) and 4U~1608-52 (Mendez et al. 1998). In the
following we discuss in turn the properties of each of the components
of the fits.

The broken power law has components similar to previous measurements
of other atoll sources. The parameters of the broken power law
indicate that that 4U~1728--34 is in the `island' state in all these
observations, i.e. at relatively low inferred mass accretion rate
(Hasinger \& van der Klis 1989).  The measurement of the break
frequency is very robust, and usually independent of the Lorentzian
components that we add to the fit.

The low-frequency Lorentzian feature at 10 to 50 Hz is a necessary
addition to the fits. Taking the February 18 interval shown in
Figure~\ref{fig:pds} as an example, we find that removing the
low-frequency Lorentzian increases the $\chi^2$/dof from 275/223 to
439/226. The probability for random variations to exceed the
corresponding F value is $4\times10^{-4}$, indicating that the
inclusion of this Lorentzian is strongly favored. In the March 1 and
February 22 to 24 observations this feature is quite wide. With a
value for $Q$ (=FWHM/$\nu$) of $1.0\pm0.2$, it does not constitute a
QPO in the usual sense.

The final component in the spectral fits are the kilohertz QPOs, which
have properties similar to those in other low mass X-ray binaries (van
der Klis 1998). Only one QPO is visible in all intervals except the 16
February observations. In the cases where there is only one QPO, we
treat it as the higher frequency QPO. The main justification for this,
is that then all the data follow one correlation of frequency versus
spectral shape (Kaaret et al. 1998). The QPOs are quite wide on March
1, with a $Q$ of about 1.

We note that there are indications of additional features in the power
spectra, for example two possible peaks on top of the low--frequency
Lorentzian in 1 Mar or 24 Feb. In these cases our adopted four
component fit function is not entirely descriptive, but the extra
features are not well constrained by the data.

\section{Results}

\begin{figure*}
\figurenum{2}
\epsscale{1.9}
\plotone{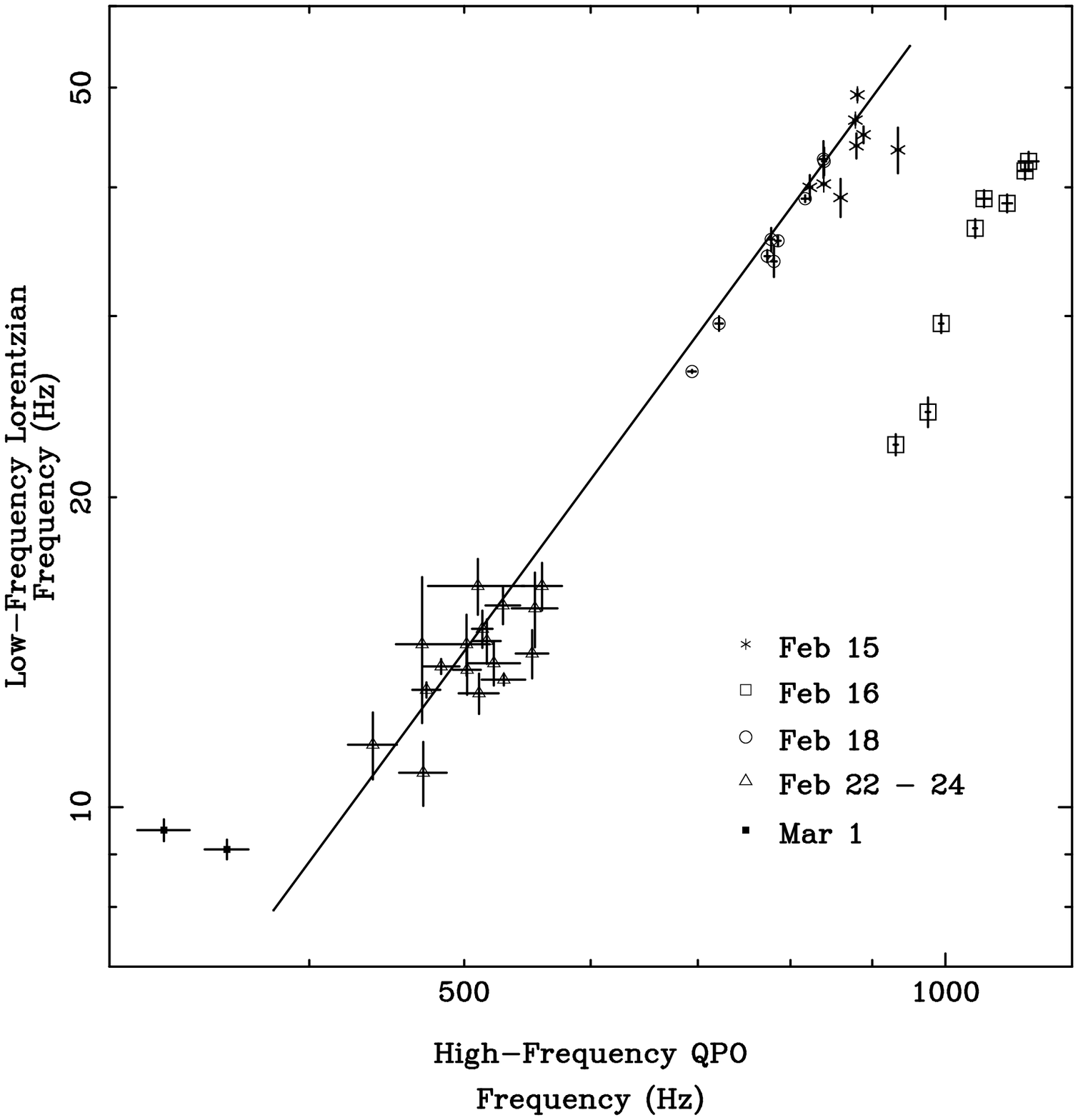} 
\caption{Frequeny of the Lorentzian at low frequencies versus the
frequency of the kilohertz QPOs. The function shown is a power law
with index 2.11. This is a fit to all of the data except 16 February
(see text).}
\label{fig:nuhi_nulo}
\end{figure*}

Figure~\ref{fig:nuhi_nulo} shows the relation between the
low--frequency Lorentzian and the kilohertz QPO. The relation between
the two frequencies can be described by a power law scaling of the
form $\nu_{LF Lor}=A\nu_{kHz}^{\alpha}$, where $\nu_{LF Lor}$ is the
frequency of the low-frequency Lorentzian and $\nu_{kHz}$ is the
centroid frequency of the kilohertz QPO.  We find $\alpha=2.11\pm
0.06$ and $A=2.9(\pm 1.1)\times10^{-5}$, excluding the 16 February
intervals (see below). The $\chi^2/{\rm dof}$ of this fit is large,
4.7, which probably is a result of not including systematic errors to
take into account the small inadequacies in the power spectral fit
function as discussed above.  The error on the power law fit is not
accurate due to the large $\chi^2$. To arrive at a more realistic
value, we multiply all data errors by a factor of two making
$\chi^2/{\rm dof}=1.2$. The quoted errors (also above) use this
procedure and $\Delta \chi^2=1$.

\begin{figure*}
\figurenum{3}
\epsscale{1.9}
\plotone{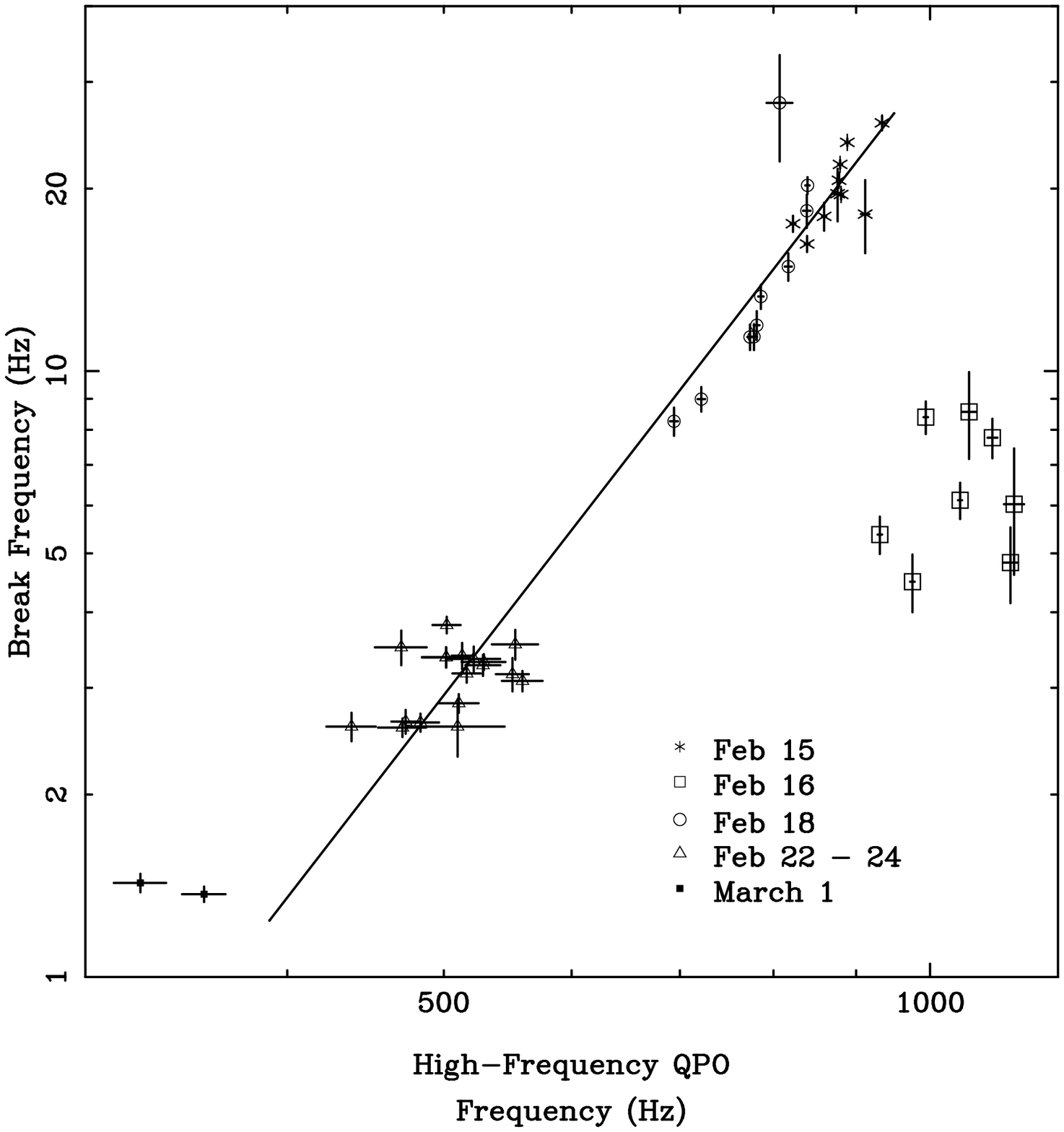}
\caption{Break frequency of the broken power law function versus the
frequency of the kilohertz QPOs. The function shown is a power law
with index 3.44, the best--fit to the data excluding 16 February.}
\label{fig:nu_break}
\end{figure*}

The second notable correlation is between the kilohertz QPOs and the
break frequency of the broken power law, as shown in
Figure~\ref{fig:nu_break}.  We describe the correlation by a function
of the form $\nu_{break}=B\nu_{kHz}^{\beta}$, where $\nu_{break}$ is
the break frequency. Using the same procedure outlined above, and
again excluding the 16 February intervals, we find $\beta=3.44\pm
0.09$ and $B=1.5(\pm 0.9)\times10^{-9}$.

As the frequency of the break increases other parameters change as
well. The level of the noise at low frequency decreases and the count
rate increases. The relation between the rms fraction and the break
frequency consistent with that found for the black hole candidates
(Mendez \& van der Klis 1997).

The only exceptions to the above correlations are in the 16 February
data set. This observation is at the highest count rate, with
presumably the highest accretion rate and highest break
frequency. However, the best--fit values for both the break frequency
and the low--frequency Lorentzian are lower than would be
expected. Perhaps these two components are not well separated in the
fits, or perhaps the Lorentzian is another component different than
any of the four components that we use in fitting the other data
(e.g. a harmonic?).  Based on the present data, it is not clear if the
16 February observations represent fundamentally different spectra. We
fit them in the same way but treat the results on somewhat different
footing.

\section{Discussion}

The previous section demonstrates a strong correlation between the
frequencies of the kilohertz QPOs and the frequencies of both the
break and the low--frequency Lorentzian, even though these features
are independently determined in the power spectra.

The low--frequency Lorentzian in this atoll source appears to be very
similar to the horizontal branch oscillation (HBO) in Z--sources.  In
particular, the correlation with the kilohertz QPOs is nearly
identical in GX17+2 (Wijnands et al. 1997), GX5-1 (Wijnands et
al. 1998) and Sco~X-1 (van der Klis et al. 1997).  This suggests that
these signals have a similar origin in both atoll and Z--sources, and
calls into question the magnetospheric beat-frequency model long used
to explain the HBOs (Alpar \& Shaham 1985).  If this model is to
account for the low--frequency signals in atoll sources, where the
mass accretion rate is roughly 100 times lower, the magnetic field
must be 10 times smaller. This is probably not consistent with the low
fields inferred for the accreting X--ray pulsar SAX J1808.4-3658
(Wijnands \& van der Klis 1998a), which is similar to atoll sources in
other respects (Wijnands \& van der Klis 1998b).

Stella \& Vietri (1998) have proposed that the low--frequency
Lorentzian feature is a result of relativistic frame dragging in the
inner accretion disk (Lense--Thirring precession). The frequency of
the slower signal correlates with the frequency of the kilohertz QPO,
assumed to be fixed by the Keplerian frequency at the inner edge of
the disk. We find a scaling index of $2.11\pm0.06$, remarkably close
to the predicted value of 2. The only free parameter in the model is
the ratio $I_{45}/M$, where $I_{45}$ is the moment of inertia in units
of $10^{45}$~g~cm$^{2}$ and $M$ is the mass of the neutron star in
units of the solar mass. We find a value for $I_{45}/M$ of 3 to
4. This is large for proposed equations of state.

Alternatively, the low frequency signal may be one of the oscillation
modes in a boundary layer in the inner disk suggested by Titarchuk,
Lapidus \& Muslimov (1998). The modes in the vertical direction are
responsible for the kilohertz QPOs by Comptonization of incoming
photons. The low frequency signals are probably not connected to modes
of rotational splitting. To get the proper scaling with the frequency
of the kilohertz QPO, the `s' parameter (Titarchuk, Lapidus \&
Muslimov 1998) would have to change by a factor of about 5. The low
frequency oscillation, however, may correspond to modes in the radial
direction. The predicted scaling is similar to the one reported here
(Titarchuk 1998).

The correlation of break frequency and kilohertz QPO frequency may be
the result of a mutual dependence on another parameter. The obvious
candidate is the mass accretion rate. Models explaining the kilohertz
QPOs predict, for different reasons, that the QPO frequency increases
with mass accretion rate (Miller, Lamb \& Psaltis 1998, Titarchuk,
Lapidus \& Muslimov 1998). It is a reasonable conjecture that the
break frequency is also correlated with the mass accretion rate, as
the power spectra are very similar to the black hole candidates where
indeed the break frequency is believed to increase with accretion rate
(van der Klis 1994a,b). This similarity is underlined by the fact that
relation between rms level of the power law and break frequency in
4U~1728--34 fits on the relation established for black holes (Mendez \&
van der Klis 1997, Belloni \& Hasinger 1990).

The question remains as to what causes the low frequency noise and
what determines the frequency of the break. The noise at low frequency
may result from a superposition of shots (e.g. Belloni \& Hasinger
1990) with the break corresponding to the shots with the longest
durations. The time scale of the shots may be set by the lifetime of
clumps (van der Klis 1994b). The clumps could be destroyed by shearing
in the inner disk.  The decreasing shearing timescale at smaller radii
(Pringle 1981) could account for the correlation of the break
frequency, mass accretion rate and kilohertz QPO frequency if the
inner disk radius decreases when the mass accretion rate increases.
Alternatively the lifetime of a shot may be determined by a diffusion
time scale in the boundary layer described in Titarchuk, Lapidus \&
Muslimov (1998). The predicted correlation of the frequencies is
similar to the one measured (Titarchuk 1998).

We thank the RXTE team for their outstanding efforts and efficiency in
making this data public. We thank Lev Titarchuk, the referee. We thank
Rudy Wijnands and Peter Jonker for interesting discussions. ECF
acknowledges support by the Netherlands Foundation for Research in
Astronomy with financial aid from the Netherlands Organization for
Scientific Research (NWO) under contract numbers 782-376-011 and
781-76-017.


\begin{figure*}
\figurenum{}
\epsscale{2.9}
\plotone{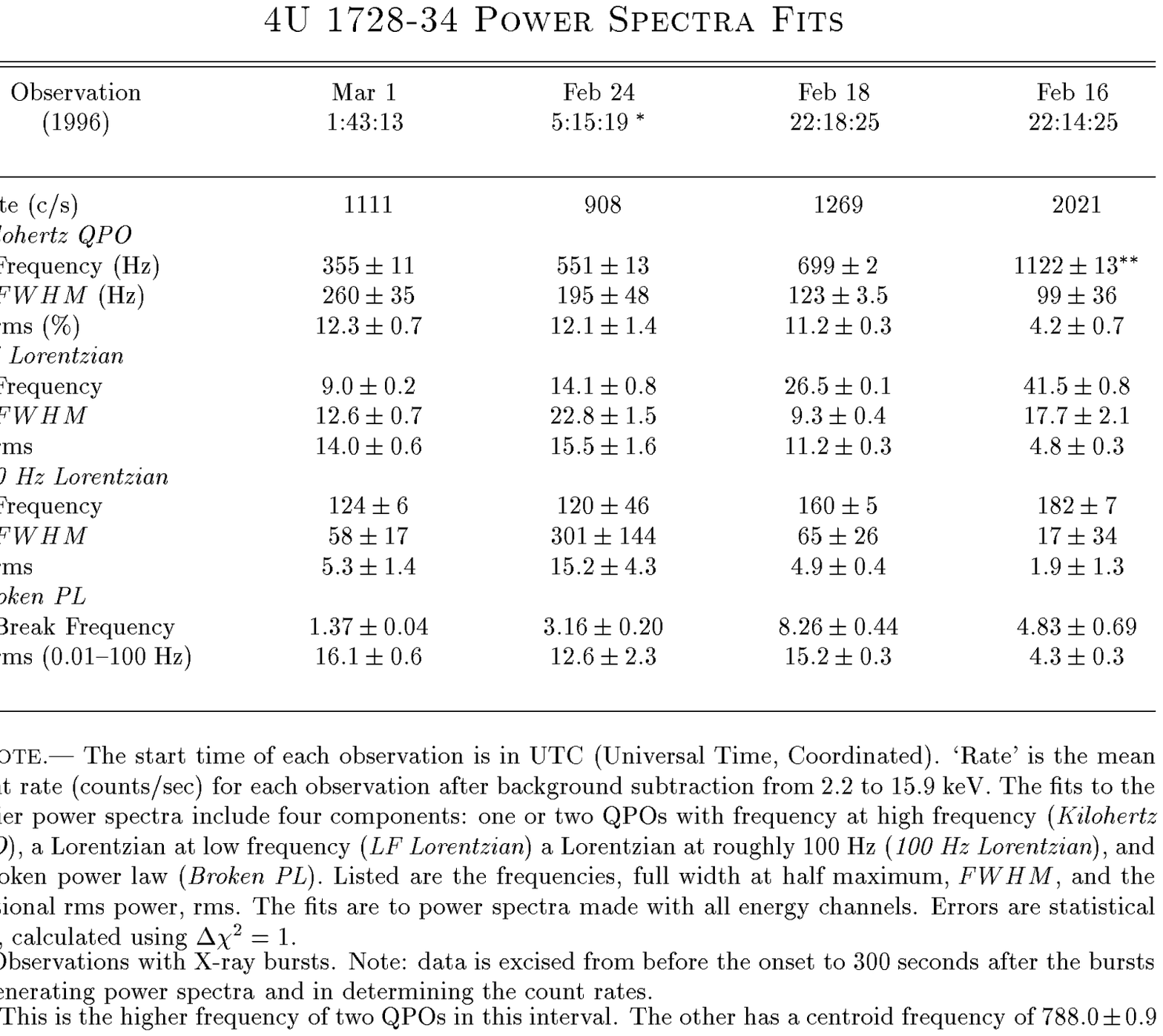} 
\caption{ }
\label{tbl:fits}
\end{figure*}


\end{document}